\documentclass[doublecol]{epl2}

\title{Polaronic Pseudogap in the Metallic Phase of La$_{0.625}$Ca$_{0.375}$MnO$_3$ Thin Films}
\shorttitle{Title} 

\author{Udai Raj Singh\inst{1} S. Chaudhuri\inst{1} R. C. Budhani\inst{1} \and Anjan K. Gupta\inst{1}}
\shortauthor{Udai Raj Singh \etal}

\institute{
  \inst{1} Department of Physics, Indian Institute of Technology Kanpur, Kanpur 208016, India.\\}


\abstract{
The electronic density of states (DOS) of La$_{0.625}$Ca$_{0.375}$MnO$_3$(LCMO) strain free epitaxial thin films with an insulator-metal transition temperature (T$_{IM}$)  of 250 K was probed using variable temperature scanning tunneling microscopy and spectroscopy. We find a depression in the DOS with a finite zero bias conductance (ZBC) signifying a pseudogap (PG) in 78-310 K temperature range. With cooling, the ZBC is found to increase indicating an increased DOS near E$_F$. We interpret PG as a signature of Jahn-Teller polarons while ZBC change, in agreement with the bulk insulator-metal transition, optical Drude peak, and photoemission experiments indicates the presence of free carriers at the Fermi energy in the metallic phase. The free carriers are discussed in terms of correlated polaronic states.}

\begin{document}

\maketitle

\section{Introduction}
The hole doped colossal magneto-resistive (CMR) manganites \cite{mang-rev} have attracted much attention in the past two decades for their intriguing physics and application potential. The Zener double exchange \cite{zener} mechanism that explained the CMR behavior was found to be insufficient \cite{millis} and this led to a more elaborate theory by Millis, Shraiman, and Mueller (MSM) incorporating Jahn-Teller (JT) interaction \cite{MSM-pol}. As evidenced by a number of experiments, like Hall effect \cite{hall-mobility}, transport \cite{pol-trans}, X-ray spectroscopy \cite{exafs}, scattering \cite{scatt-pol}, and isotope effect \cite{isotope}, the JT small polarons (SP) \cite{holstein} seem to be responsible for the activated resistivity in the paramagnetic insulating (PI) phase of manganites. Several experiments such as neutron scattering \cite{scatt-pol}, optical conductivity \cite{optical-pol}, photoemission \cite{manella-PE-PRB}, and tunneling \cite{tunn-renner, tunn-fischer}, have also shown signatures of polarons in the ferromagnetic metallic (FM) phase; although these signatures are weaker, at least the structural ones \cite{scatt-pol, exafs}, than those in PI phase. Moreover, the optical Drude weight \cite{optical-pol} and recent angle resolved photoemission spectroscopy (ARPES) \cite{ARPES-QP} observation of a quasiparticle peak at E$_F$ indicate the presence of free carriers as well in the FM phase.

Besides MSM theory \cite{MSM-pol}, which finds a transformation of SP into delocalized carriers at T$_{IM}$, a number of other models have also been proposed in the recent past. Emin \cite{emin} suggested that these delocalized carriers are large polarons with larger spatial extent of lattice deformation. Alexandrov and co-workers \cite{alexandrov} advocate the splitting of singlet bipolarons at high temperatures into SP below T$_{IM}$. Both these models propose only polaronic states in the FM phase. Another simple model, put forward by Ramakrishna and coworkers \cite{TVR-lb-model}, accounts for the free carriers by having delocalized band states together with localized polarons. In addition, this model proposes a ``coherent SP" state at low temperatures to account for small magnitude of resistivity. Therefore, both the theory and experiment point towards some kind of polaron softening or delocalization with the onset of the ferromagnetic order. However, the detailed physics of the FM phase seems far from understood. In particular, the role of charge carriers, i.e. whether they are free or polaronic, in the FM state is not yet clear.

The phase separation (PS) scenario \cite{PS-rev}, i.e.  two types (SP and delocalized) of coexisting carriers, is also consistent with a large number of experiments. PS has been established experimentally in some of the narrow bandwidth manganites \cite{PS-exp-NB}. In wide bandwidth manganites some tunneling experiments do indicate electronic inhomogeneities but mostly in granular or inhomogeneous samples \cite{PS-exp-WB} whereas the homogeneous samples do not seem to show any PS \cite{AKR-PRB, tunn-fischer}.

In this paper we report the variable temperature Scanning Tunneling Microscopy and Spectroscopy (STM and STS) studies of strain free epitaxial La$_{0.625}$Ca$_{0.375}$MnO$_3$ (LCMO) thin films with T$_{IM}$= 250 K. The tunneling spectra show a polaronic pseudogap. The STS images show some electronic inhomogeneities corresponding to small variations in zero bias conductance (ZBC). We also see an increase in the PG energy at T$_{IM}$ with cooling and a rise in the ZBC signifying the build-up of states at the Fermi energy with cooling. This observation, consistent with the bulk resistivity, optical and recent ARPES data, strongly supports a correlated polaron state giving rise to delocalized carriers in the metallic state.

\section{Experimental Details}
Epitaxial La$_{0.625}$Ca$_{0.375}$MnO$_3$ (LCMO) films of thickness 200 nm were grown on NdGaO$_{3}$ (NGO) (110) substrates using pulsed Laser deposition. The samples, mounted on the STM sample holder with a conducting silver epoxy, were transferred into the STM cryostat in a very short time ($<$30min) to minimize the time of exposure to air. The STM cryostat was evacuated to high vacuum ($\sim$10$^{-4}$ mbar) before cooling. Tunneling spectra and the STS images were acquired using ac-modulation technique; however for ZBC comparison direct I-V spectra were used. For the spectra at a fixed point the feedback was switched off while for the STS imaging the feedback was kept on but with a larger time constant so as it does not respond to the ac-modulation. In this way one gets a contrast in STS images which actually anticorrelates with the metallicity, i.e., the darker regions have more DOS at E$_{F}$ while the brighter ones have less DOS at E$_{F}$ . We kept the junction resistance values same for all the local spectra taken at different locations and different temperatures to minimize the variation in the tip-sample separation. This is necessary for comparing the absolute values of dI/dV for different spectra at a particular bias voltage. A dI/dV-V spectrum is a convolution of the density of states (DOS) and the energy dependent matrix element, which can be normalized away \cite{tun-norm-ref} by plotting the normalized conductance (dI/dV)/(I/V), i.e. dlnI/dlnV. However, at V=0, dlnI/dlnV=1 by definition, so for studying the variation in DOS at E$_F$ ZBC(dI/dV at V = 0 ) is analyzed.

\section{Results and Analysis}
\begin{figure}
\onefigure[width=7cm,height=6cm]{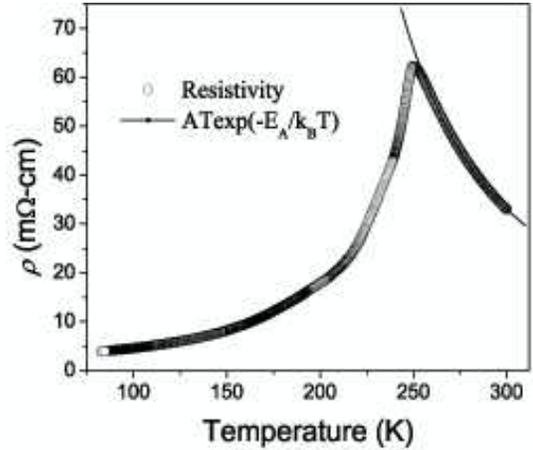}
\caption{The four probe resistivity of the LCMO thin film with T$_{IM}$=250K. The continuous line is the fitting with polaronic activation behavior with activation energy E$_A$=0.112eV.}
\label{fig:resistivity}
\end{figure}

Fig.~\ref{fig:resistivity} shows the four probe resistivity as a function of temperature for an LCMO thin film with an insulator to metal transition at 250 K. Above 250 K the film shows insulating behavior with a polaron activation gap of E$_A$=0.112 eV (or 1310 K) as shown by the fitting $\rho=ATexp(-E_A/k_BT)$ \cite{holstein} in fig.~\ref{fig:resistivity}. The absolute resistivity is in m$\Omega$-cm range and the decrease across T$_{IM}$ in resistivity is by a factor of two. This change is not so large as compared to some of the narrow bandwidth manganites\cite{udai}. Thus we would expect lesser change in DOS near E$_{F}$ across T$_{IM}$ as compared to the narrow bandwidth manganites.

\begin{figure}
\onefigure[width=4cm,height=3.7cm]{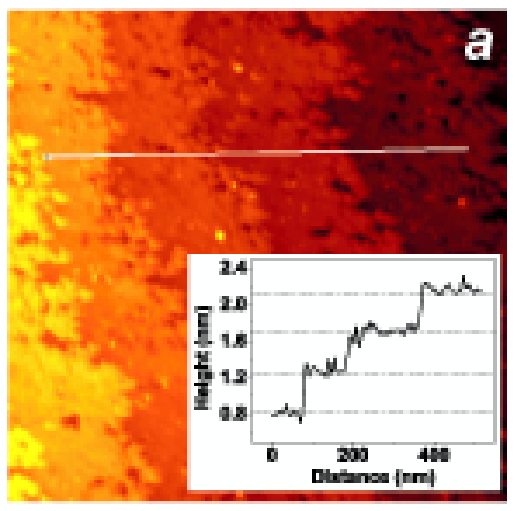}
\onefigure[width=4cm,height=3.7cm]{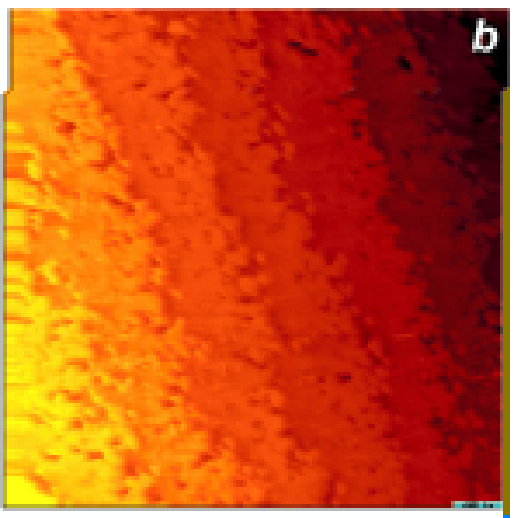}
\caption{ Topographic image of LCMO film at {\bf a} 310 K (area 601$\times$601 nm$^{2}$, bias = 1.0 V, current = 0.1 nA) and {\bf b} 78 K (748 $\times$ 748 nm$^{2}$ , 1.0 V/0.1 nA). The inset in {\bf a} shows the line cut with atomic height terraces.}
\label{fig:topography-LCMO}
\end{figure}

\begin{figure}
\onefigure[width=4cm,height=3.5cm]{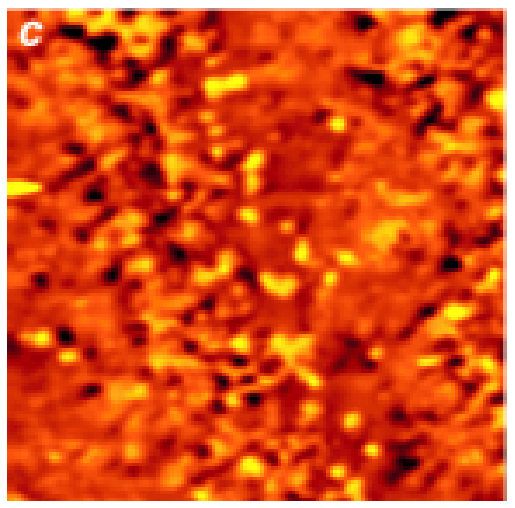}
\onefigure[width=4cm,height=3.5cm]{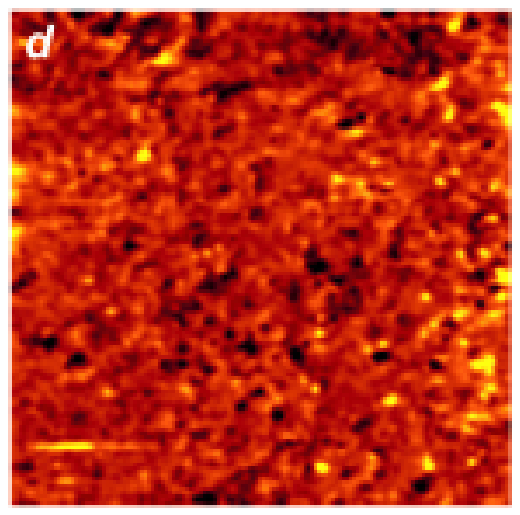}
\caption{{\bf a} and {\bf b} show the simultaneous topographic and conductance image (area 304 $\times$ 304 nm$^{2}$, 1 V/0.1 nA) taken at 78 K. The film surface have been found electronically homogeneous at all temperatures except some insulating bright regions of 5-10nm$^{2}$ area.}
\label{fig:conductance-LCMO}
\end{figure}

Fig.~\ref{fig:topography-LCMO} shows the topographic STM images of this film at 310 and 78 K showing the atomic terraces of 0.4($\pm$0.05) nm step height and 100-150 nm width. Each terrace has a roughness of about 0.1 nm. The terraces are observed over more than 2$\times$2 $\mu$m$^2$ area at all temperatures between 78 and 310 K. The STS images (see fig.~\ref{fig:conductance-LCMO}b) show little contrast which consists of some isolated bright spots with a weak contrast in the background. The bright spots show large gap spectra while in other places we see pseudogapped spectra with a small variation in the ZBC that correlates well with the STS image contrast. We believe that these insulating bright spots in STS images correspond to some local chemical defects \cite{udai}.

\begin{figure}
\onefigure[width=8.5cm,height=4cm]{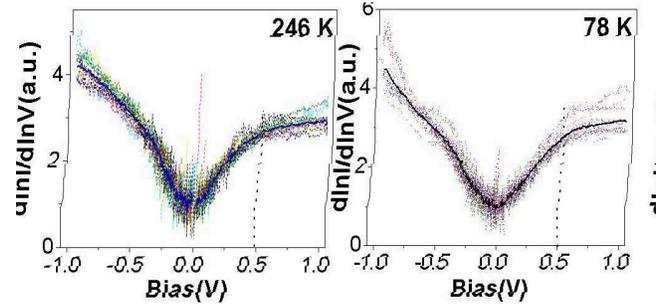}
\caption{ The dotted curve show spectra those were taken at different places. The solid line curve in figure 246 K and 78 K  show average spectra.}
\label{fig:average_spctra}
\end{figure}

\begin{figure}
\onefigure[width=7cm,height=6cm]{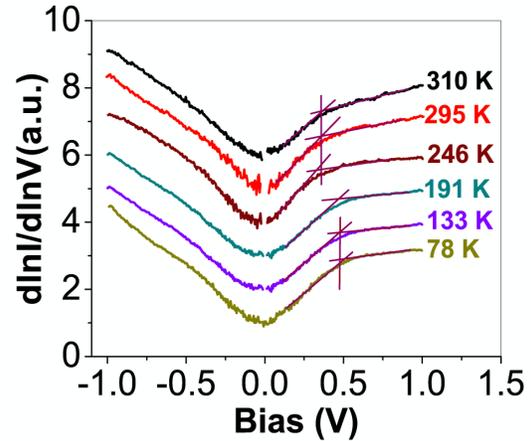}
\caption{The temperature dependent spectra taken at 10G$\Omega$ (1 V/ 0.1nA) junction resistance. dlnI/dlnV spectra have been offset uniformly for clarity.}
\label{fig:norm_cond_spectra}
\end{figure}

Fig.~\ref{fig:average_spctra} shows the normalized spectra at 246 and 78 K taken at different locations in a particular area together with their average. These spectra are neither metal-like nor pure gap-like and they feature a PG like depression near zero bias with a finite ZBC. Although the spectra at these two temperatures look qualitatively similar but there is a difference in the energy scale of the PG which is quite clear despite the spatial variations in the spectra. To eliminate such variations we have taken an average of several tens of spectra at different places to study the evolution of the DOS with temperature. The corresponding dlnI/dlnV spectra are shown in fig.~\ref{fig:norm_cond_spectra}.  We see that the spectra at all temperatures are quite asymmetric. Such asymmetry is quite common in most transition metal oxides including cuprates \cite{cupr-tunn-asym} and its origin is not quite clear. It could also arise from the effects related to tunneling matrix element as argued by Ukraintsev \cite{ukraintsev}, particularly for large bias range, which is the case here. We see that for negative bias the kink due to gap-like feature in the normalized conductance is still visible but much weaker because of a rising background.

The temperature evolution of the normalized spectra across T$_{IM}$ can be described by two features. The first one is the energy-scale over which the depression in DOS (representing PG) occurs, which can be qualitatively deduced from the spectra as the voltage at which there is a change in slope. This can be seen better with the guiding lines corresponding to the low bias and high bias part of the spectra as shown in fig.~\ref{fig:norm_cond_spectra}. Thus the pseudogap's energy-scale increases with cooling by about 0.2V as can be seen by comparing the spectra above and below 246 K. The second feature is the decrease in the slope of the pseudogap edge across T$_{IM}$ with cooling. This can be seen from the guidelines of low bias region in the same figure. Thus with cooling the gap edge does not really move out to increase the pseudogap but it becomes wider.

\begin{figure}
\onefigure[width=7.2cm,height=5.8cm]{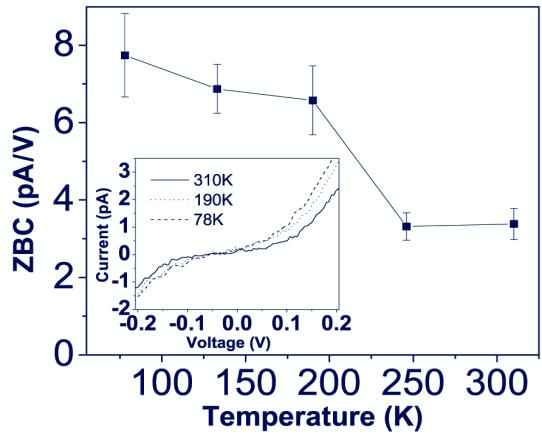}
\caption{The variation in ZBC with temperature with the error bars showing the rms spread in ZBC. The inset shows representative low bias region of the I-V spectra at three different temperatures taken at same junction resistances (1V / 0.3nA).}
\label{fig:gap-var}
\end{figure}

The other major result of our measurements is a non-zero DOS at E$_F$ i.e., the ZBC value is finite at all temperatures and it increases with cooling. We see a clear variation in the ZBC with temperature as seen in figure \ref{fig:gap-var}; the zoomed in plot of the average I-V spectra are also shown in the inset of figure \ref{fig:gap-var}. All the I-V spectra used in this plot have been taken with the same junction resistance (1 V/0.3 nA) and each spectra is an average of several tens of spectra taken at different locations but avoiding the insulating spots. We clearly see that the slope of the I-V spectra near zero bias increases with cooling and in particular, it has a jump near T$_{IM}$.  This increase in ZBC with cooling is more in-line with the bulk transport measurements, low temperature optical Drude weight \cite{optical-pol}, and the recent ARPES results \cite{ARPES-QP}.

The above two results, i.e. increase in ZBC and PG energy scale with cooling, are indicative of some spectral weight transfer from high energy (roughly from 0.35 eV to 0.50 eV) to the lower energies. The PG does not disappear in FM state of manganite which is unusual for the metallic phase of manganites. The PG has also been observed by optical, ARPES and recent tunneling measurements \cite{optical-pol,Chikamatsu-Saitoh,Mitra,tunn-renner,tunn-fischer} and it is an indication of JT polarons.

\section{Discussions}
Tunneling measures the DOS, which, in comparison to ARPES, represents the spectral function integrated over all {\bf k} vectors. The recent ARPES experiments \cite{ARPES-QP} find pockets of quasiparticles in a very narrow region of the 2-D Fermi surface of a layered manganite. Thus the change in the total DOS at E$_F$ due to this quasiparticle peak would be rather small. The earlier ARPES on bilayer manganites did not observe this peak \cite{no-QP-ARPES} due to an insufficient {\bf k}-resolution. We believe that these quasiparticles at E$_F$ exist in the metallic phase of other manganites as well but require much better {\bf k} resolution to be experimentally detected. The 2-D bilayer manganites offer an ideal system for ARPES as they provide excellent {\bf k}-resolution. The c-axis tunneling may not be able to probe the in-plane quasiparticles for 2-D manganites \cite{tunn-renner} as the in-plane {\bf k}-states contribute negligible in c-axis tunneling. In 3-D manganites the same {\bf k} resolution is difficult with ARPES since well oriented surfaces, like 2-D cleavable manganites, are not possible to prepare.

The MSM theory \cite{MSM-pol}, for a range of JT coupling, finds delocalized states in the FM phase giving a non-zero spectral weight or DOS at E$_F$. The measured Drude weight \cite{optical-pol} is two orders of magnitude smaller than in a typical metal, although it does indicate the presence of metal-like free carriers in FM phase. However, the exact nature of the mobile charge carriers in FM state is far from clear. In particular, it is still debatable whether manganites are bad metals because they are homogeneous systems with all (polaronic) carriers having poor mobility or they are a mixture of free carriers and localized polarons.

It seems from various experimental results, including the present one, that we can neither rule out free carriers nor polarons in FM phase. Thus the coexistence of two types of carriers is a likely scenario. The delocalized carriers in FM phase could arise in a number of ways, such as large polarons (LP), band electrons, or coherent SP. Although our tunneling data cannot differentiate between these scenarios but we believe that the small polarons' coherence \cite{TVR-lb-model, manella-PE-PRB, Zhao-pol} is giving rise to the metallic free carriers, particularly in the broad bandwidth manganites.

We would like to point out that in most manganites, the surface has been found to be less conducting with a weaker ferromagnetic order \cite{surf-mag-weak,udai}. The breaking of crystal symmetry or oxygen deficiency at film surface could be affecting the surface's electronic and magnetic properties\cite{Choi}. However, the fact that we see a change in PG behavior at T$_{IM}$ and the temperature variation of ZBC qualitatively agrees with the bulk transport makes us believe that the surface here does reflect the bulk, at least, qualitatively. It is also possible that the surface provides additional barrier (other than the vacuum barrier) for tunneling into the bulk.

In summary, our tunneling experiment finds a PG at all temperatures (78-310K) in the strain free epitaxial LCMO thin films. The PG energy is larger in the FM phase but the DOS at E$_F$ is also larger in FM state than the PI state. The latter result is in agreement with the bulk transport, optical conductivity, and recent ARPES results. From this we strongly believe that the IM transition in the broad bandwidth manganites is resulting from a polaronic transition involving the high temperature SP giving rise to coherent and delocalized states at E$_F$.

We would like to acknowledge financial support from the DST of the Govt. of India. URS acknowledges CSIR for financial support.

\end{document}